\documentclass[%
 reprint,
superscriptaddress,
 amsmath,amssymb,
 aps,
pra
]{revtex4-2}
\usepackage{graphicx} 
\usepackage{gensymb}
\usepackage{svg}
\usepackage{xcolor}

\begin{document}
\title{Quantum emitters in bilayer hexagonal boron nitride}
\author{Reyhan Mehta}
\affiliation{Department of Condensed Matter Physics and Materials Science, Tata Institute of Fundamental Research, Homi Bhabha Road, Mumbai 400005, India}
\author{Anshuman Kumar}
\email{anshuman.kumar@iitb.ac.in}
\affiliation{%
Laboratory of Optics of Quantum Materials, Department of Physics, Indian Institute of Technology Bombay, Mumbai- 400076, India 
}%
\affiliation{Centre of Excellence in Quantum Information, Computation, Science and Technology, Indian Institute of Technology Bombay, Powai, Mumbai 400076, India}
\date{\today}

\begin{abstract}
Hexagonal boron nitride has been experimentally shown to exhibit room-temperature single-photon emission. This emission is attributed to defect states in the wide band-gap of hBN, which allow new optical transitions between these dispersion-less defect levels. In this work, we study the new spectral features introduced by interacting atomic defects in consecutive layers of bilayer hBN. Density Functional theory simulations have been carried out to calculate the energy band structure, frequency-dependent complex dielectric functions, and Kohn-Sham states to demonstrate and understand the cause of the emission enhancements. We found that placing colour centres in the vicinity of each other in bilayer hBN introduces new polarization dependent spectral features, with strong dependence on the distance and relative orientation between atomic defects. Our results provide a pathway to engineering single photon emission in hBN via inter-defect interaction. 
\end{abstract}


\maketitle

\section{Introduction}
\begin{figure}[h]
    \centering
    \includegraphics[width=1.0\linewidth]{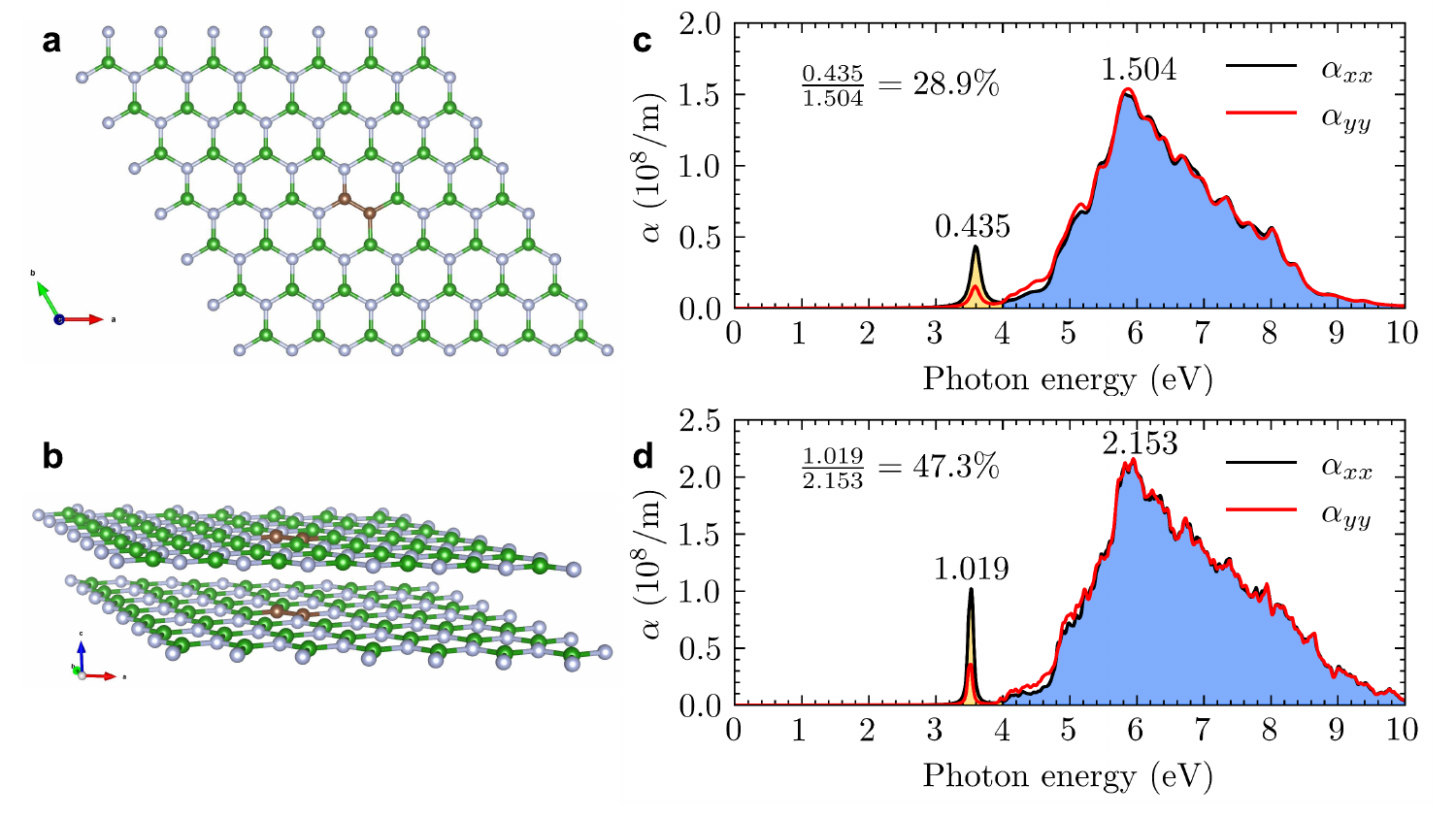}
    \caption{\textbf{Emission enhancement in defected bilayer hexagonal boron nitride.} \textbf{a} and \textbf{b} show the schematic diagrams for monolayer and bilayer hBN (with C$_\text{B}$C$_\text{N}$ point-defect), respectively. \textbf{c} and \textbf{d} show the optical absorbance spectra for the systems in \textbf{a} and \textbf{b}, respectively. The labelled numbers are the peak heights for the Lorentzian region (yellow) and bulk region (blue) of $\alpha_{xx}$.}
    \label{fig1}
\end{figure}
Defect engineering in hexagonal boron nitride is a promising pathway towards achieving single photon emission in the ultraviolet, visible, and near-infrared spectral regions\cite{tran2016quantum, doi:10.1021/acsnano.6b03602, PhysRevApplied.5.034005, PhysRevB.106.014107, C7NR04270A}. Theoretical studies on colour centres in hBN traditionally investigate the electronic and optical properties of atomic point-defects in monolayer hBN. The energy and brightness of these emissions is determined by the nature of the defect and its interaction with the host lattice (the monolayer hBN). In this work, we study the new spectral features introduced by placing colour centres in the vicinity of each other in bilayer hBN. In addition to this, our study also proposes a method to tune the emission energies, by altering the relative orientation and distance between the defects in the two layers.

Although studies have been carried out to investigate the effect of combining individual atomic defects in monolayer hBN \cite{PhysRevB.106.014107}, extensive research has not been conducted yet to investigate the tuning of the quantum emission by changing the relative distance and orientation between individual defects. Combining defects in monolayer hBN has the effect of eliminating spin-splitting and introducing a single new defect state above the Fermi energy (but below the bulk conduction bands), leading to a strong optical transition between the two defect states (marked as 1 and 2 in Fig. \ref{fig6}b). However, when such a defect dimer is introduced in both layers of bilayer hBN, the number of total defect states is doubled, with a slight energy difference between the electronic levels (the states marked as 1, 2, 3, and 4 in Fig. \ref{fig7} b). The new feature offered by defects in bilayer hBN is that by tuning the distance and orientation between the bilayer dimer defects, we can break or restore inversion symmetry of the defect state wavefunctions, which imposes certain selection rules determining the allowed optical transitions between defect states in bilayer hBN. This allows us to dictate how many optical transitions are allowed between the multiple defect states, and in some cases also allows us to create a polarized quantum emitter (see Fig. \ref{fig8}). In addition to this, combining defects in bilayer hBN also leads to enhancement in the single photon emission.

The enhancement in spectral features is quantified by the relative heights of the peaks, corresponding to the single photon emission and bulk-band transitions, in the absorbance spectrum (see Fig.~\ref{fig1}). The absorbance in Fig.\ref{fig1} is calculated from the frequency-dependent complex dielectric function using the following formula \cite{hung2022quantum}:
\begin{equation}
    \alpha(\omega) = \frac{2\omega}{c}\sqrt{\frac{|\epsilon(\omega)| - \text{Re}[\epsilon(\omega)]}{2}}.
\end{equation}
In Fig. \ref{fig1}, the absorbance spectra for monolayer and bilayer hBN show that there is an enhancement of 18.4\% in the relative peak heights (the peak height of the Lorentzian region divided by the peak height of the bulk region, as shown in Fig. \ref{fig1}c and Fig. \ref{fig1}d), when two C$_\text{B}$C$_\text{N}$ defects are allowed to couple with each other in bilayer hBN. The enhancement is explained by the splitting of the bands in the band structure, when a second layer of hBN (with C$_\text{B}$C$_\text{N}$ defect) is superposed on top of the first layer of hBN (also with C$_\text{B}$C$_\text{N}$ defect). The band splitting gives rise to extra defect states below and above the Fermi energy, allowing for extra optical transitions with nearly the same energies. This result suggests that the emission from the bilayer is not simply an incoherent sum of the two monolayer defect spectra but rather shows a complicated behaviour involving inter-defect interaction. In the following sections, we study this interaction in more details highlighting the tunability of the spectral features of the single photon emission in the bilayer system. We begin by studying the optical and electronic properties of pristine monolayer and bilayer hBN, and see how these properties change with the addition of C$_\text{B}$C$_\text{N}$ defects in the monolayer and bilayer systems. Furthermore, we also investigate how the complex dielectric tensor changes when the distance and orientation between the defects is varied in bilayer hBN, and we explain these changes by looking at the Kohn-Sham wavefunctions for the resulting defect states.

\section{Pristine monolayer and bilayer hBN}
\begin{figure*}[h]
    \centering
    \includegraphics[width=0.9\linewidth]{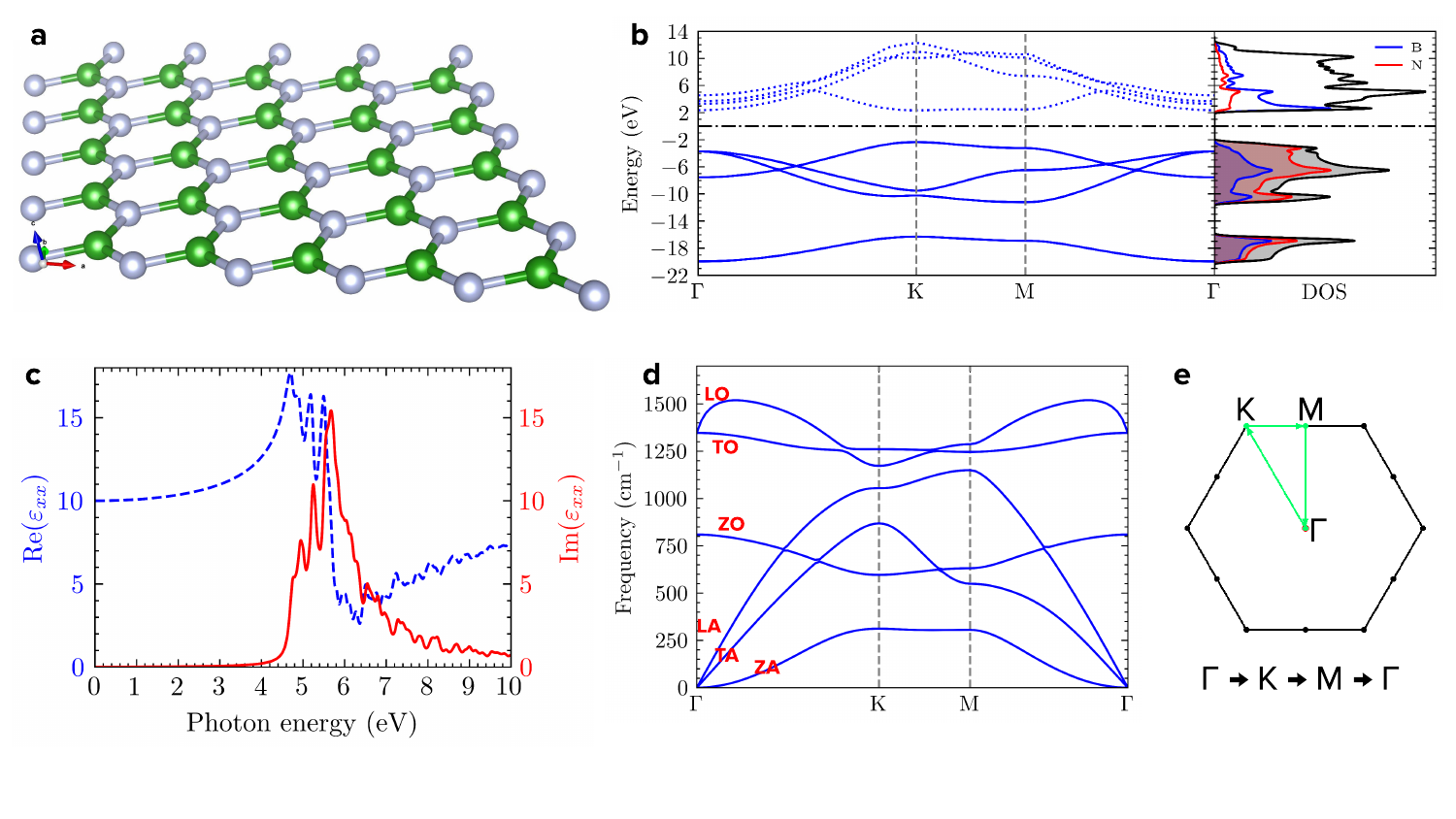}
    \caption{\textbf{Properties of pristine monolayer hBN}. \textbf{a,} Schematic illustration of hBN monolayer. \textbf{b,} Electronic energy dispersion and density of states of monolayer hBN. \textbf{c,} Frequency dependent complex dielectric function of monolayer hBN. \textbf{d,} Phonon dispersion of monolayer hBN. \textbf{e,} Brillouin zone of hBN monolayer.}
    \label{fig2}
\end{figure*}
\begin{figure*}[h!]
    \centering
    \includegraphics[width=0.9\linewidth]{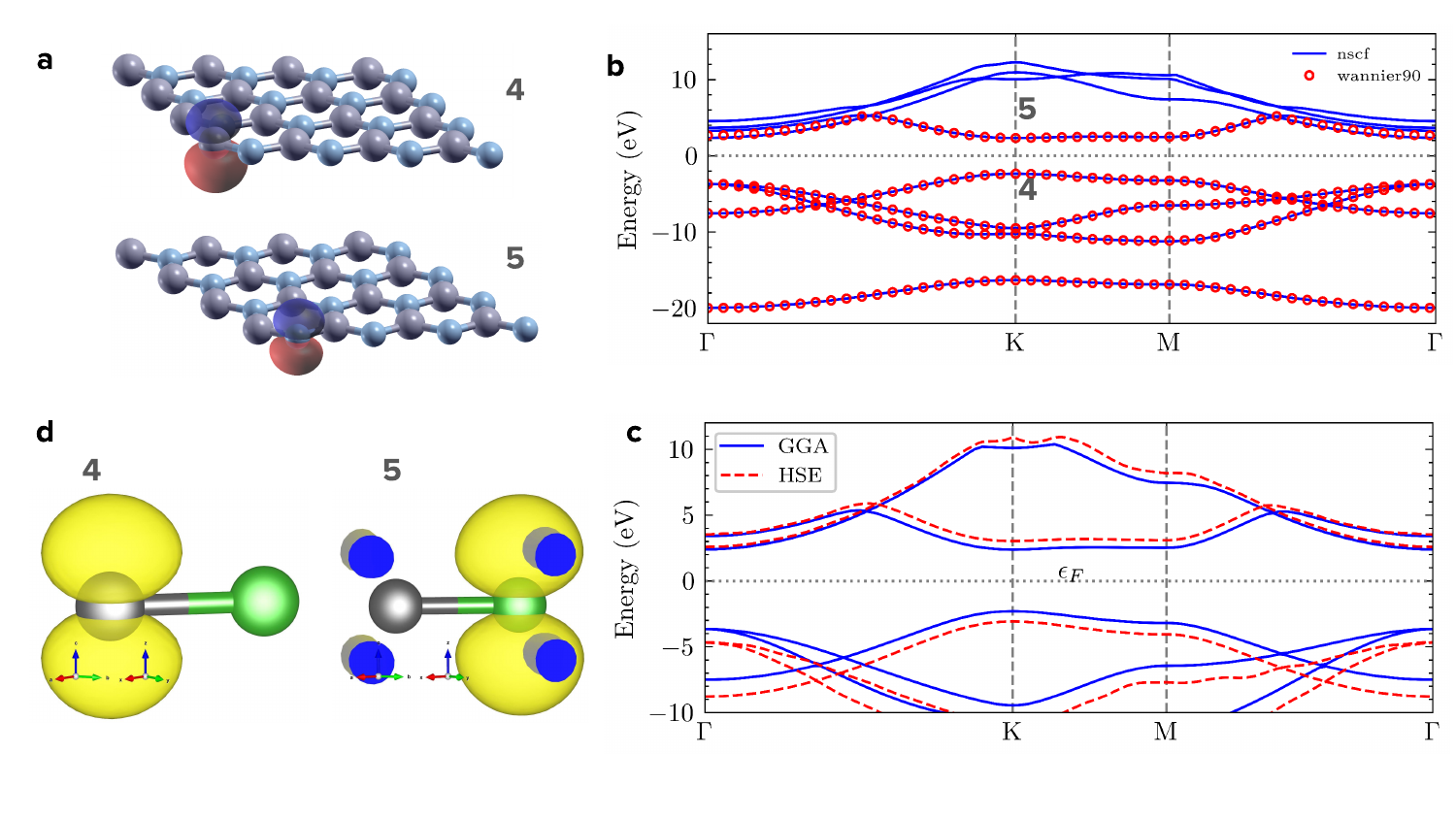}
    \caption{\textbf{Wavefunctions of pristine monolayer hBN.} \textbf{a,} Selected maximally-localized Wannier functions of monolayer hBN, marked in GGA band structure. \textbf{b,} GGA band structure of monolayer hBN using Wannier interpolation and NSCF calculation. \textbf{c,} Hybrid functional band structure of monolayer hBN, using Wannier interpolation. \textbf{d,} Selected Kohn-Sham wavefunctions at K-point for bands marked in GGA band structure.}
    \label{fig3}
\end{figure*}
\begin{figure*}[h]
    \centering
    \includegraphics[width=0.9\linewidth]{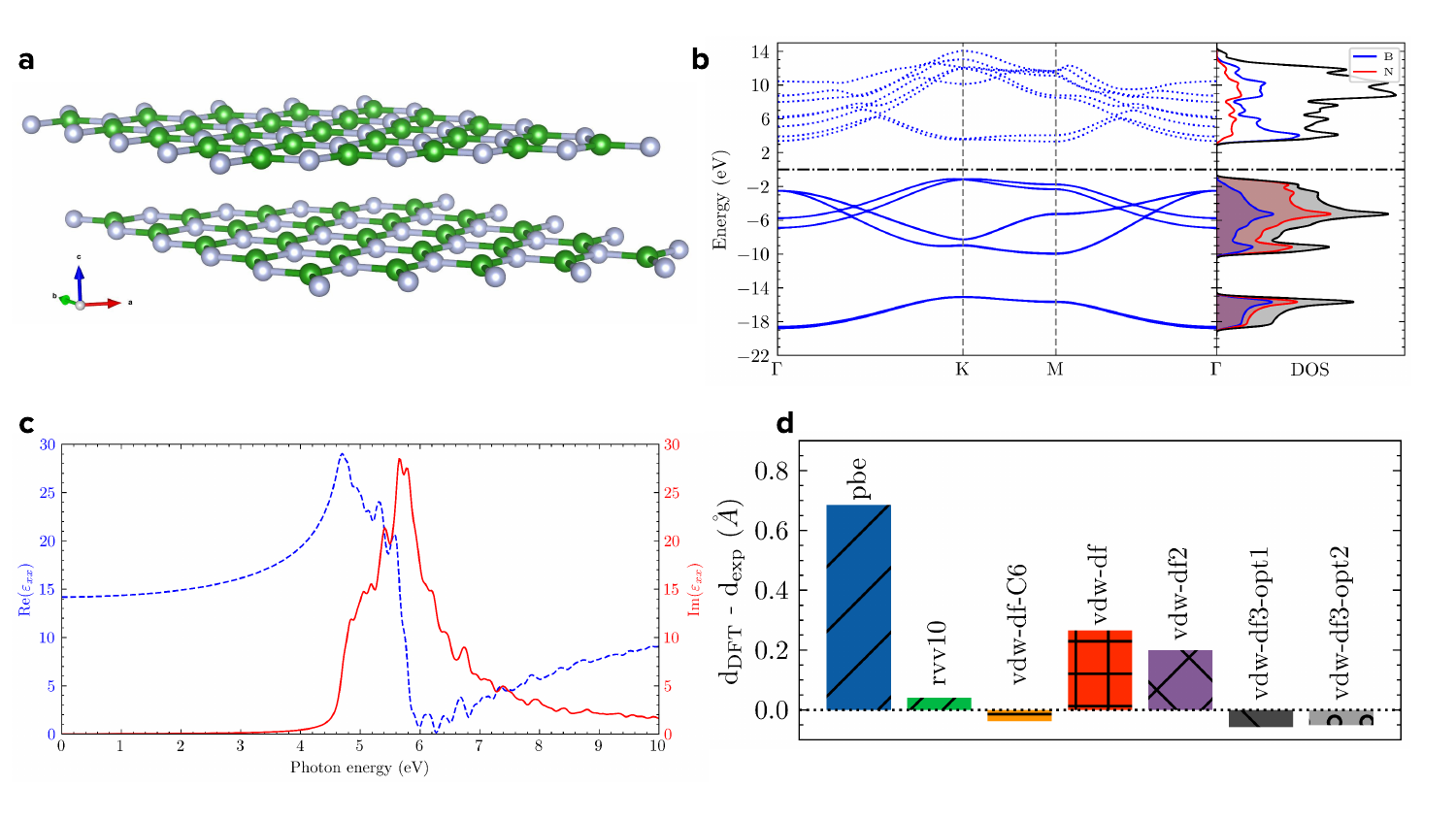}
    \caption{\textbf{Properties of pristine bilayer hBN (AA’ stacking).} \textbf{a,} Schematic illustration of hBN bilayer. \textbf{b,} Electronic energy dispersion and density of states of bilayer hBN. \textbf{c,} Frequency-dependent complex dielectric function of bilayer hBN. \textbf{d,} Comparison of various non-local van der Waals’ functionals for bilayer hBN.}
    \label{fig4}
\end{figure*}
\begin{figure*}[h!]
    \centering
    \includegraphics[width=0.9\linewidth]{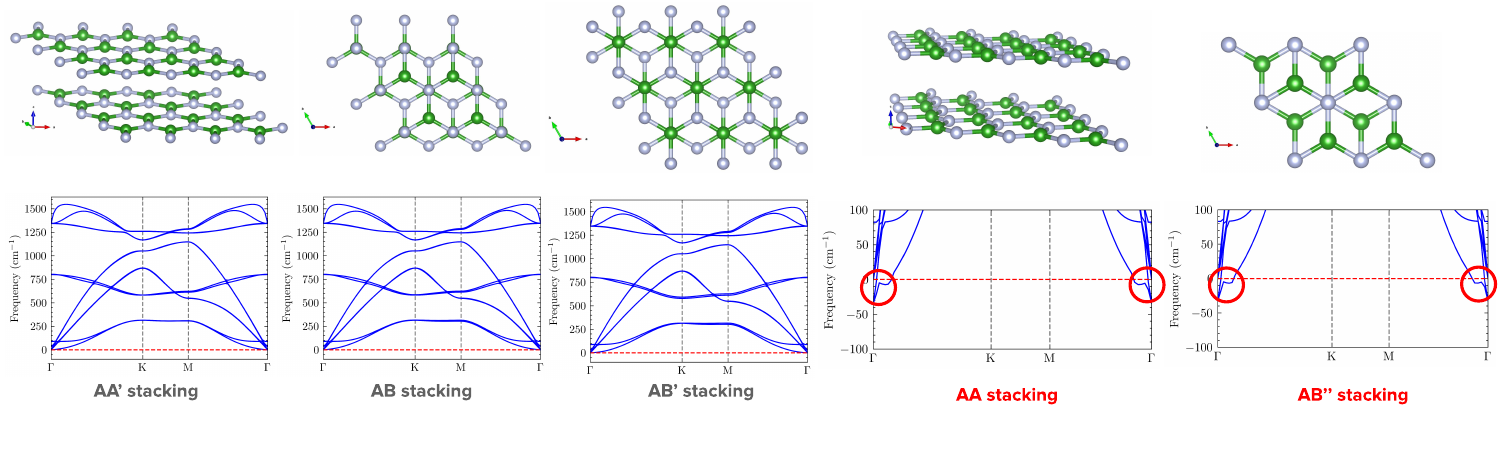}
    \caption{\textbf{Phonon dispersion for stable and unstable stacking sequences of bilayer hBN.} The phonon eigenfrequencies for the stable configurations are always non-negative, indicating stability. For the unstable configurations, imaginary modes (indicated by negative frequencies) appear at the $\Gamma$-point for AA and AB'' stacking sequences in bilayer hBN.}
    \label{fig5}
\end{figure*}
This study is initiated by an analysis of the electronic and optical properties of \textit{pristine} monolayer and bilayer hexagonal boron nitride. The schematic illustration of monolayer hBN is shown in Figure \ref{fig2}a. Structurally, hBN is identical to graphene (hence the alternative name \textit{white graphene} for hBN), with the B-N bond lengths $d = 1.45\mathring{\text{A}}$, as predicted by these DFT calculations, and also by experiment \cite{doi:10.1021/acs.chemmater.5b03607}. The electronic energy dispersion and the atom-resolved projected density of states for monolayer hBN was also calculated, and is plotted in Figure \ref{fig2}b. The Brillouin-zone path, which connects high-symmetry points in the reciprocal space of monolayer hBN is described in Figure \ref{fig2}e \cite{KOKALJ1999176},\cite{kokalj2000scientific}. It can be seen from the band structure that the DFT calculation with the GGA functional predicts a direct band-gap (at the K-point) of about 4.6 eV. The projected density of states suggests that the highest occupied valence orbital in monolayer hBN is localized at a Nitrogen site, and that the lowest unoccupied conduction orbital is localized at a Boron site. This is confirmed by Figure \ref{fig3}: The Maximally-localized Wannier functions in Figure \ref{fig3}a and the Kohn-Sham states (at the K-point) in Figure \ref{fig3}d (both labelled in Figure \ref{fig3}b), verify that indeed the valence orbital is p$_z$-like, localized at the Nitrogen atom, and the conduction orbital is also p$_z$-like, but localized at the Boron atom. It is thus the electronic transition between the p$_z$ orbitals of the Nitrogen and Boron atoms which defines the band-gap of pristine monolayer hBN. Incidentally, the hybrid-functional band structure in Figure \ref{fig3}c suggests that monolayer hBN has a direct band-gap of about 6 eV at the K-point, which is consistent with other first-principles studies carried out for hBN \cite{doi:10.1021/acs.jpcc.8b09087}, and it also verifies that although the GGA DFT calculations underestimate the band-gap (as expected), the deviation is not too high. Henceforth, all DFT calculations for this study were done using a GGA functional (in combination with a non-local vdW functional).

The frequency-dependent complex dielectric function in Figure \ref{fig2}c shows that the imaginary part of $\varepsilon_{\text{xx}}$ starts rising at about 4.6 eV, which is consistent with the calculated band structure in Figure \ref{fig2}b. The region after 4.6 eV corresponds to the wide band-gap transitions in hBN, and is expected to appear for defected hBN as well (as can be seen in Fig. \ref{fig6}c).

The phonon dispersion in Figure \ref{fig2}d shows the phonon bandstructure of pristine monolayer hBN. At $\Gamma \rightarrow 0$, these can be classified as: LA, TA, and ZA (Longitudinal, Transverse, and Out-of-plane acoustic modes); and LO, TO, and ZO (Longitudinal, Transverse, and Out-of-plane optical modes). These are labelled in Figure \ref{fig2}d, and were determined by inspecting the displacement eigenvectors corresponding to each mode. It can also be seen in Figure \ref{fig2}d that because hBN is a 2D polar material, the phonon dispersion is non-analytic at the $\Gamma$-point, and there is no LO-TO splitting; this is consistent with past experimental data \cite{li2024observation}.

The electronic energy dispersion, atom-resolved projected density of states, and frequency-dependent complex dielectric function were also computed for pristine \textit{bilayer} hBN (AA' stacked). It can be seen from the band structure in Figure \ref{fig4}b that the bands split away from the K-point; this is supposedly due to the interaction between the two layers of hBN. This same splitting is also expected to appear for defect states, as can be verified by Fig. \ref{fig6}b and Fig. \ref{fig7}b. The dielectric function in Figure \ref{fig4}c illustrates that the optical absorption spectrum of bilayer hBN is similar to that of monolayer hBN, with the exception of smoothing of the sharp features, owing to the fact that more than one closely-spaced optical transitions exist in bilayer hBN, due to the splitting of the bands.

A geometry relaxation was carried out for pristine bilayer hBN using a variety of non-local vdW functionals, and the interlayer spacing was compared for each functional, to the experimental value; the result is plotted in Figure \ref{fig4}d. The experimental value for the interlayer distance $c = 3.315\mathring{\text{A}}$ is taken to be the representative value for bulk hBN \cite{doi:10.1021/ct200880m}. According to these DFT calculations, Figure \ref{fig4}d suggests that the \textbf{vdw-df-C6} functional is best at predicting the interlayer distance of bilayer hBN; henceforth this vdW non-local functional was used for this study. Furthermore, only the AA' and AB stacked sequences are considered in this study; certain other stacking sequences represent unstable phases of bilayer hexagonal boron nitride, as indicated by the negative frequencies in their corresponding phonon dispersion graphs, as shown in Fig. \ref{fig5}. This is also confirmed by calculation of the phonon dispersion for the same systems, using the frozen phonon method, in another study \cite{li2024exceptionally}. It can be seen in Figure \ref{fig5} that there is a splitting of the phonon bands away from the K-point; this is reminiscent of the band splitting in Figure \ref{fig4}b.

\section{C$_\text{B}$C$_\text{N}$ defect in hBN}
\begin{figure*}[h]
    \centering
    \includegraphics[width=0.9\linewidth]{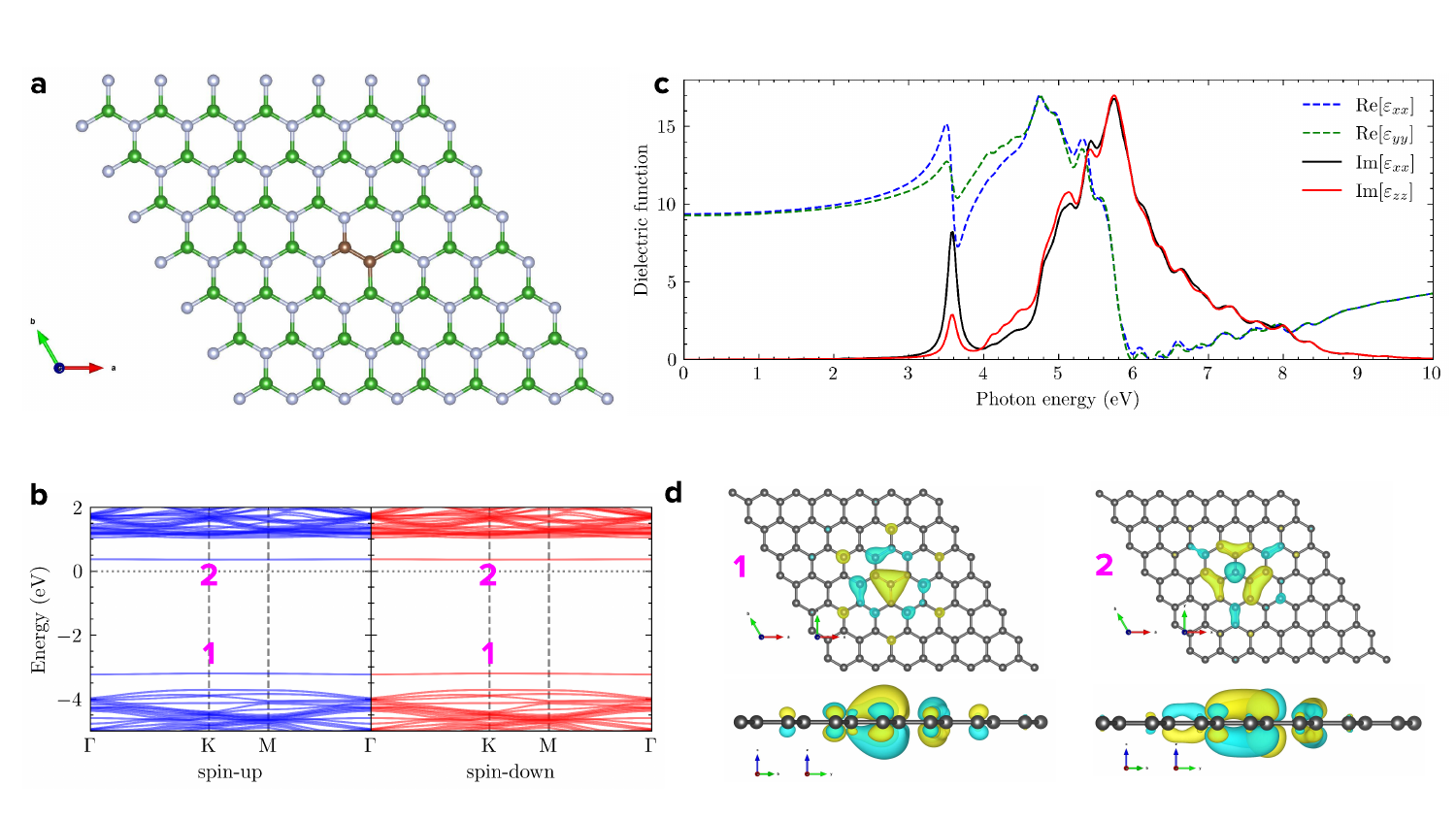}
    \caption{\textbf{Properties of C$_\text{B}$C$\text{N}$ defect in monolayer hBN.} \textbf{a,} Schematic illustration of hBN monolayer with C$_\text{B}$C$_\text{N}$ defect. \textbf{b,} Spin-resolved electronic energy dispersion of C$_\text{B}$C$_\text{N}$ defect. \textbf{c,} Frequency dependent complex dielectric function of C$_\text{B}$C$_\text{N}$ defect. \textbf{d,} Kohn-Sham wavefunctions at the $\Gamma$-point, of defect states marked in band structure.}
    \label{fig6}
\end{figure*}
\begin{figure*}[h]
    \centering
    \includegraphics[width=0.9\linewidth]{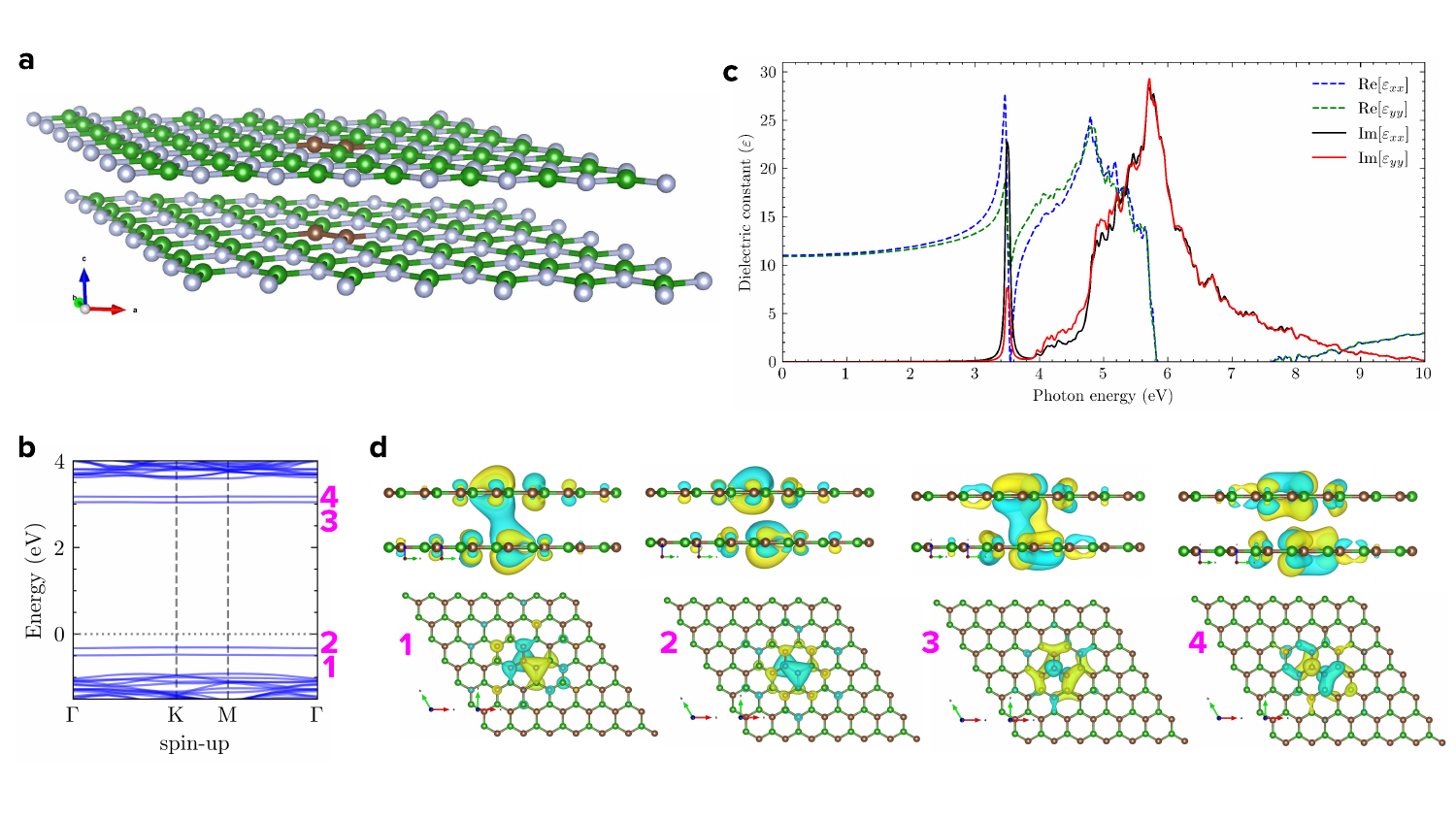}
    \caption{\textbf{Properties of C$_\text{B}$C$_\text{N}$ defect in bilayer hBN.} \textbf{a,} Schematic illustration of hBN bilayer with C$_\text{B}$C$_\text{N}$ defect. \textbf{b,} Spin-resolved electronic energy dispersion of C$_\text{B}$C$_\text{N}$ defect. \textbf{c,} Frequency dependent complex dielectric function of C$_\text{B}$C$_\text{N}$ defect. \textbf{d,} Kohn-Sham wavefunctions at  $\Gamma$-point, of defect states marked in band structure.}
    \label{fig7}
\end{figure*}

DFT calculations were performed for the C$_\text{B}$C$_\text{N}$ defect in monolayer and bilayer hBN, and the resulting band structures, dielectric functions, and Kohn-Sham orbitals are plotted in Fig. \ref{fig6} and Fig. \ref{fig7}. The imaginary part of the complex dielectric function has a peak at 3.5 eV, and this is consistent with the energy difference between the occupied and unoccupied defect states marked as 1 and 2 in Figure \ref{fig6}b. The C$_\text{B}$C$_\text{N}$ defect can be seen as a combination of the C$_\text{B}$ and C$_\text{N}$ defects; the C$_\text{B}$C$_\text{N}$ dimer acts like a simple molecule embedded within an hBN lattice, with its energy levels defined by the $\pi$-bonding between the p$_\text{z}$ orbitals of the Carbon atoms at the Nitrogen and Boron sites \cite{PhysRevB.106.014107}. The spin-splitting is absent for the C$_\text{B}$C$_\text{N}$ defect, because the electronic structure resembles the Highest Occupied Molecular Orbital (HOMO, state marked as 1 in Figure \ref{fig6}b) and Lowest Unoccupied Molecular Orbital (LUMO, state marked as 2 in Figure \ref{fig6}b) of a molecule; since the HOMO level is fully occupied and LUMO level is empty, the spin-splitting is absent\cite{PhysRevB.106.014107}. In fact, it was found that spin-splitting is absent for all DFT calculations performed for this study, and thus only spin-up band structures and electronic energy levels are shown henceforth.

In a similar fashion, DFT calculations were also performed for bilayer hBN, with the C$_\text{B}$C$_\text{N}$ defects lying face-to-face on each of the two separate layers; the results are shown in Figure \ref{fig7}. As can be seen in Figure \ref{fig7}c, the Im[$\epsilon_{\text{xx}}$] (black curve) has a taller and sharper peak at 3.5 eV, in comparison to that for monolayer hBN with C$_\text{B}$C$_\text{N}$ defect (Figure \ref{fig6}c). This enhancement is only present for the $xx$-component of the dielectric tensor.

The imaginary part of the complex dielectric tensor, $\text{Im}[\epsilon_\text{xx}]$, can be seen as a response function derived from perturbation theory with adiabatic turning on (from Quantum ESPRESSO's internal manual for $\texttt{epsilon.x}$):
\begin{multline}
    \text{Im}[\epsilon_\text{xx}] = 1 + \frac{4\pi e^2}{\Omega N_\textbf{k}m^2} \sum_{n,n'} \sum_{\textbf{k}}\frac{{\textbf{M}_{xx}}}{(E_{\textbf{k},n'}-E_{\textbf{k},n})^2} \\\left\{ \frac{f(E_{\textbf{k},n})}{E_{\textbf{k},n'}-E_{\textbf{k},n} + \hbar \omega +i\hbar \Gamma} + \right. \\ \left. \frac{f(E_{\textbf{k},n})}{E_{\textbf{k},n'}-E_{\textbf{k},n} - \hbar \omega -i\hbar \Gamma} \right\},
\end{multline}
where $\Gamma$ is the adiabatic parameter, and for the conservation of total energy $\Gamma \rightarrow 0$, which transforms the fractions in this equation to Dirac Delta functions, implying that every excited state has an infinite lifetime. In the limit of small $\Gamma$, the expression for the dielectric tensor is given by the Drude-Lorentz model.

\begin{figure*}[h]
    \centering
    \includegraphics[width=0.9\linewidth]{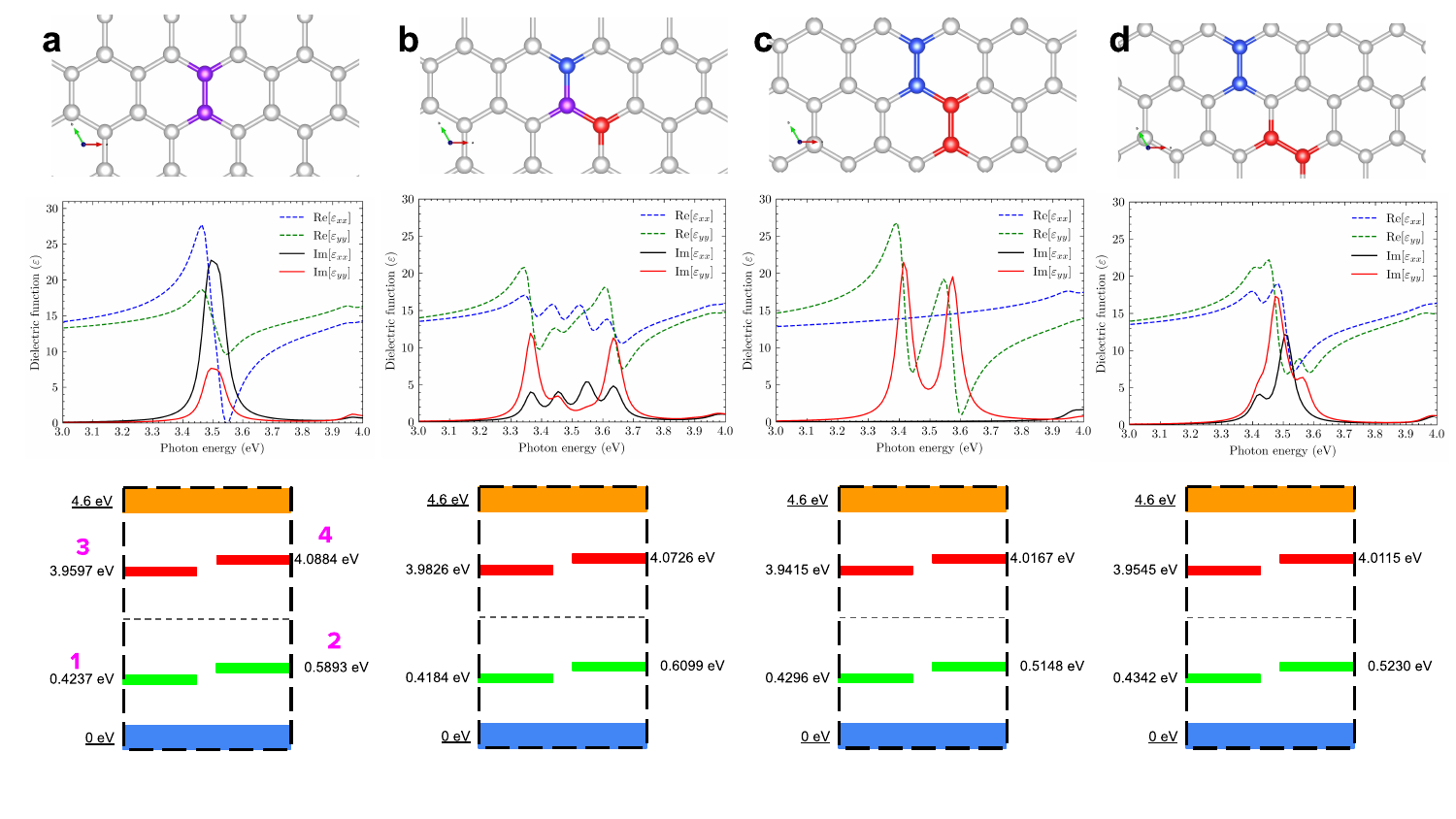}
    \caption{\textbf{Dependence of the dielectric function on the relative distance between bilayer defects.} The top row indicates how the defects are placed relative to each other; the blue atoms denote carbon atoms in the upper hBN layer, and the red atoms denote carbon atoms in the lower hBN layer. The purple atoms denote carbon atoms in both the upper and lower hBN layers. For example, Fig. \ref{fig8}a denotes the same system in Fig. \ref{fig7}a.}
    \label{fig8}
\end{figure*}

\begin{figure*}[hbt!]
    \centering
    \includegraphics[width=0.9\linewidth]{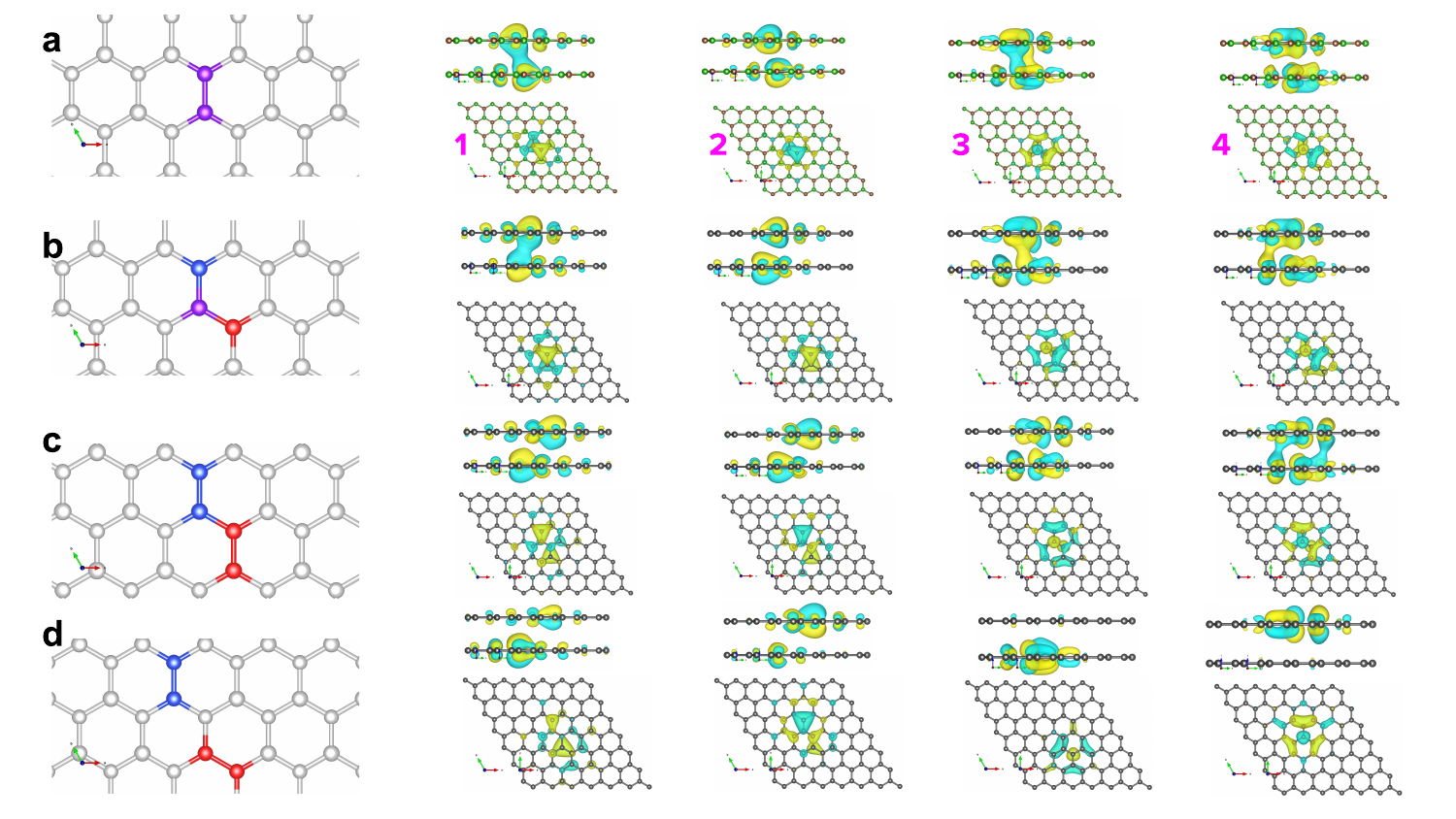}
    \caption{\textbf{The Kohn-Sham wavefunctions at the $\Gamma$-point for the bilayer systems shown in Fig. \ref{fig8}.} The colour of the orbital denotes the sign of the wavefunction. The top rows contain side-views, and bottom rows contain top-views.}
    \label{fig9}
\end{figure*}

The transition dipole moment, $\textbf{M}$, is calculated by using the formula:
\begin{equation}\label{tdm}
    \textbf{M} = \int \psi_{n'}^*(\textbf{r}) [q\textbf{r}] \psi_{n}(\textbf{r})d^3\textbf{r},
\end{equation}
where $q$ is the electron's charge ($-e$), $\textbf{r}$ is its position, and the integral is over all space. The parity of the wavefunctions can be used to explain the allowed and forbidden transitions in bilayer hBN with C$_\text{B}$C$_\text{N}$ defects; the electronic energy levels and dielectric function for various configurations of interlayer defects are shown in Fig. \ref{fig8}, and their respective Kohn-Sham wavefunctions at the $\Gamma$-point are shown in Fig.~\ref{fig9}.  


In Fig. \ref{fig8}a, it can be seen from the dielectric spectrum that only the transitions between states $1 \longleftrightarrow 3$ and $2 \longleftrightarrow 4$ are allowed. This is explained by noticing the parities of the orbitals plotted in Fig. \ref{fig9}a. This is in contrast to the system in Fig. \ref{fig8}b - all four transitions are allowed, because the inversion symmetry of the wavefunctions for the defect states is broken, and none of the integrals defined by equation (\ref{tdm}) disappear. However, for the system in Fig. \ref{fig8}c, the transitions $2 \longleftrightarrow 4$ and $1 \longleftrightarrow 3$ are \textit{forbidden}, because of parity (see Fig. \ref{fig9}c). It should also be noted from Fig. \ref{fig8}c that the defect is polarized; Im[$\epsilon_{xx}$] = 0 for the interval where Im[$\epsilon_{yy}$] $\neq 0$. For the system in Fig. \ref{fig8}d, the inversion symmetry is again broken (see Fig. \ref{fig9}d), so all four transitions are allowed. 

\begin{figure*}[hbt!]
    \centering
    \includegraphics[width=0.9\linewidth]{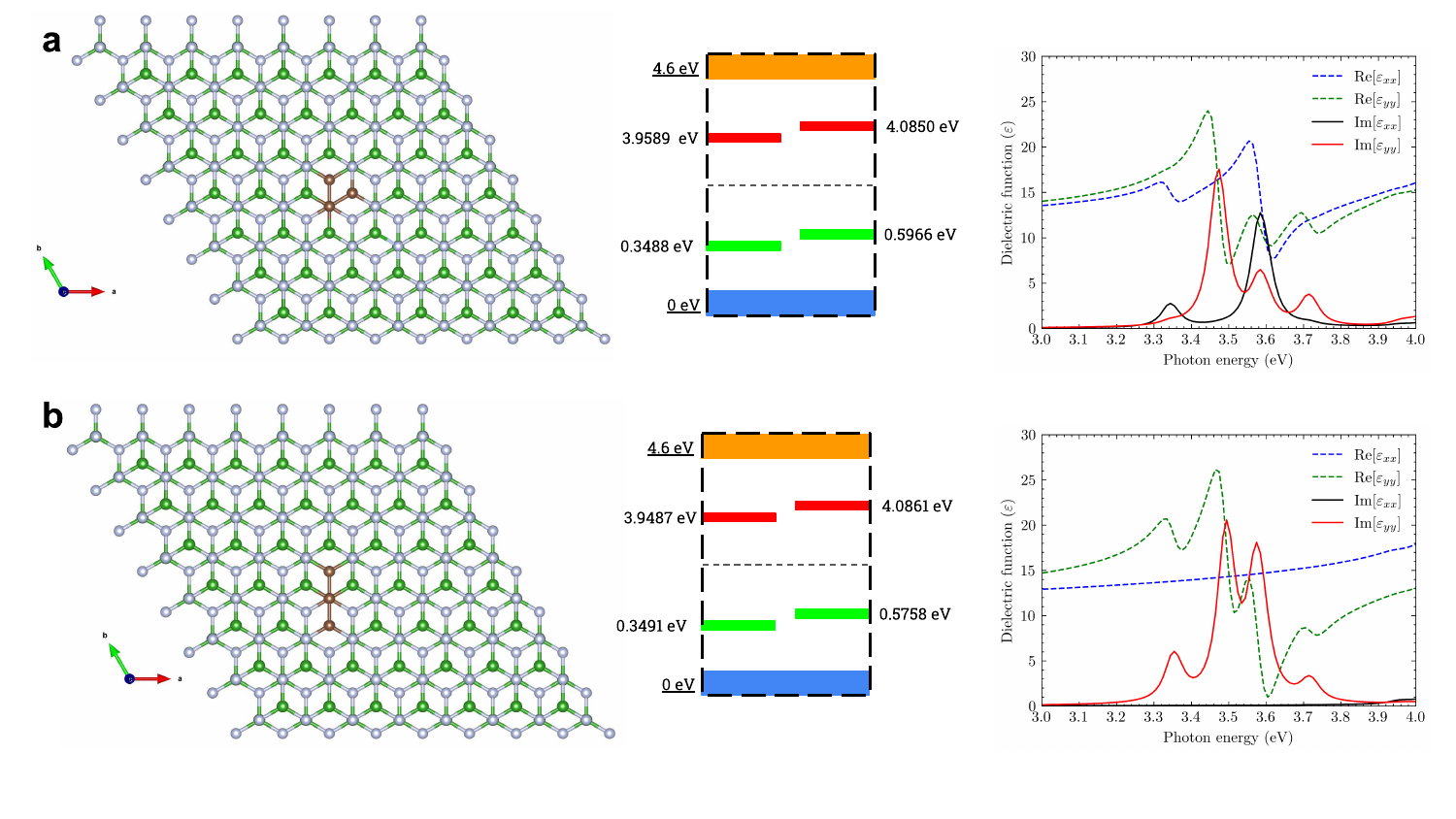}
    \caption{\textbf{The electronic energy levels and frequency-dependent complex dielectric functions for two C$_\text{B}$C$_\text{N}$ defect configurations in AB stacked bilayer hBN.} For both the configurations, all four transitions between states $1, 2 \longleftrightarrow 3, 4$ are allowed.}
    \label{fig10}
\end{figure*}

DFT calculations were also performed for AB stacked bilayer hBN with C$_\text{B}$C$_\text{N}$ interlayer defects. AB stacked bilayer hBN can be seen as a special case of twisted bilayer hBN, with a twist angle $\varphi = 60\degree$. It can be deduced from the dielectric functions in Fig. \ref{fig10} that all four transitions between states $1, 2 \longleftrightarrow 3, 4$ are allowed, which is again explained based on selection rules. It should also be noted that for the system in Fig. \ref{fig10}b, the defects in both layers are aligned along the $y-$axis, and the optical absorbance of the defect inherits this polarization (Im[$\epsilon_{xx}$] = 0). This is similar to the polarized defect in the system shown in Fig. \ref{fig8}c.

\section{Conclusion}
In summary, we performed a detailed investigation of the spectral characteristics of coupled atomic defects in bilayer hBN. It was observed that by adjusting the relative distance and orientation between the defects in the two layers, one can selectively tailor the optical selections. Finally, we report interesting polarization dependent characteristics in the optical spectra of bilayer defected hBN, with a strong dependence on the stacking sequence, distance and orientation between the bilayer defects. Our study offers a route to design single photon emission in hBN crystals, with polarization and frequency control.

\section*{Acknowledgements} 
A.K. acknowledges funding from the Department of Science and Technology via the grants: SB/S2/RJN-110/2017, ECR/2018/001485 and DST/NM/NS-2018/49.

\section{Methods}
Density Functional Theory (DFT) calculations of point defects in hBN were carried out in order to calculate the electronic energy dispersion, frequency-dependent complex dielectric tensor, and the Kohn-Sham wavefunctions of the periodic systems considered in this study. The calculations were performed using Quantum ESPRESSO \cite{QE-2017, QE-2009}, using the generalized gradient approximation (GGA) of Perdew, Burke and Ernzerhof (PBE) \cite{PhysRevLett.77.3865}. A supercell composed of 7 $\times$ 7 unit cells of hBN was used for both the monolayer (98 atoms) and bilayer (196 atoms) calculations, to reduce interactions between neighbouring periodic images of defect sites. Geometry relaxation was performed for obtaining the ground state electronic structure, using a 3 $\times$ 3 $\times$ 1 $k$-grid.

Norm-conserving pseudopotentials were used for the DFT calculations \cite{VANSETTEN201839}, \cite{PhysRevB.88.085117}. The kinetic energy cut-off for the wavefunctions was fixed at 85 Ry, and the convergence threshold for self-consistency was fixed at $10^{-9}$ Ry. For the structural energy minimization the internal coordinates are allowed to relax until the atomic forces are less than $10^{-4}$ a.u. and the total energy changes less than $10^{-5}$ a.u. between two consecutive SCF steps. The height of the supercell was fixed at 20$\mathring{\text{A}}$ to reduce interactions between vertically periodic images. The lattice constant of the hBN primitive cell was found to be $a = 2.511 \mathring{\text{A}}$, calculated on a 9 $\times$ 9 $\times$ 1 $k$-grid. This is consistent with other DFT calculations performed for hBN, using SIESTA and VASP \cite{C7NR04270A}.

After using DFT to calculate the electronic energy dispersion of these defects, the frequency-dependent complex dielectric function was calculated using the Independent Particle Approximation (IPA), which is the simplest approximation used to describe single-particle excitation. In addition to this, the Kohn-Sham wavefunctions ($|\psi(\text{\textbf{r}})|^2 \cdot \text{sign}[\psi(\text{\textbf{r}})]$) were calculated at the $\Gamma$-point for selected electronic energy states, labelled in corresponding band structure diagrams. Throughout this study, we have drawn conclusions about the transition dipole moment (defined by equation (2)) for the various defect states considered in this study, by inspecting the plots for $|\psi(\text{\textbf{r}})|^2 \cdot \text{sign}[\psi(\text{\textbf{r}})]$.

For the bilayer hBN calculations, non-local Van der Waals (vdW) functionals were used for taking into account the vdW interactions between the two layers in bilayer hBN \cite{PhysRevLett.115.136402, PhysRevB.76.125112, Berland_2015, Langreth_2009}. These functionals were first used for optimizing the structure of bilayer hBN \cite{Sabatini_2012} and then for calculating the electronic and optical properties of the same system.

The phonon dispersion in pristine monolayer and bilayer hBN was calculated using Density Functional Perturbation Theory (DFPT) \cite{RevModPhys.73.515, PhysRevB.43.7231, PhysRevLett.58.1861}. The energy convergence threshold for the plane-wave self-consistency field (SCF) step was set to $10^{-12}$ Ry, and for the phonon calculation it was set to $10^{-15}$ Ry. An 8 $\times$ 8 $\times$ 1 automatic $k$-grid was used for the SCF step, and a 6 $\times$ 6 $\times$ 1 uniform $q$-grid was used for the phonon calculation. Since hBN is a polar material, the dielectric tensor and Born effective charges were also calculated in the phonon calculation. This is because for polar materials in the $\textbf{q} \rightarrow \textbf{0}$ limit, a macroscopic field appears as a result of the long-range Coulomb interaction, which is incompatible with periodic boundary conditions; for this reason, a non-analytic term must be added to the Interatomic Force Constants at $\textbf{q} = \textbf{0}$ \cite{PhysRevB.66.235415}. This non-analytic term depends on the Born effective charges and the dielectric tensor, and they can all be calculated from the (mixed) second-order derivatives of the total energy. The system was explicitly treated as a 2D system, in order to accurately capture the (absence of) LO-TO splitting at the $\Gamma$-point for hBN \cite{doi:10.1021/acs.nanolett.7b01090}, \cite{PhysRevB.96.075448}. The acoustic sum rule (ASR) was also imposed to ensure that the frequencies of the acoustic modes at the $\Gamma$-point be zero, for stable configurations of hBN.

For the pristine monolayer and bilayer hBN calculations, the projected density of states was calculated using the following procedure: An SCF calculation was performed on a $9 \times 9 \times 1$ automatic $k$-grid, followed by a non-self-consistent field (NSCF) calculation on an automatic $36 \times 36 \times 1$ $k$-grid, with tetrahedral occupations \cite{PhysRevB.49.16223} for the Brillouin-zone integration. For the post-processing step, a simple Gaussian broadening of 0.02 Ry was used.

The Wannier functions for pristine monolayer hBN were calculated after performing an SCF calculation for the same (with a vertical vacuum separation of 15$\mathring{\text{A}}$). The initial projections for the orbitals were: three sp$^2$ and one p$_z$ orbital for Boron, all centered on the Boron, and three sp$^2$ and one p$_z$ orbital for Nitrogen, all centered on the Nitrogen. 20 iterations were used to produce the Maximally-localized Wannier functions (using the Wannier90 code \cite{Pizzi_2020}), which are in good agreement with the band structure calculated using an NSCF calculation (see Fig. \ref{fig3}b).

The Hybrid-functional band structure was calculated using the following procedure: An SCF calculation was performed on a dense, $16 \times 16 \times 1$ automatic $k$-grid, using the Heyd-Scuseria-Ernzerhof (HSE) exchange-correlation functional \cite{heyd2003hybrid}, using a $q$-grid of $12 \times 12 \times 1$ for the Fock operator. The EXX fraction for the hybrid functional calculation was chosen to be 0.25. After the SCF calculation, the Wannier functions were calculated using the same procedure as described above (but with 50 iterations for the localization of the Wannier functions), to obtain the Wannier-interpolated hybrid-functional band structure for pristine monolayer hBN (shown in Fig. \ref{fig3}d).

\bibliography{references}
\end{document}